# Magnetic Proximity Evoked Colossal Bulk Photovoltaics in Crystalline Symmetric Layers

Xingchi Mu, Qianqian Xue, Yan Sun, Jian Zhou[*]

*Center for Alloy Innovation and Design, State Key Laboratory for Mechanical Behavior of Materials, Xi'an Jiaotong University, Xi'an, 710049, China*

Bulk photovoltaic (BPV) effect, a second order nonlinear process that generates static current under light irradiation, requires centrosymmetric broken systems as its application platform. In order to realize measurable BPV photocurrent in spatially centrosymmetric materials, various schemes such as chemical doping, structural deformation, or electric bias have been developed. In the current work, we suggest that magnetic proximity effect via van der Waals interfacial interaction, a contact-free strategy, also breaks the centrosymmetry and generate large BPV photocurrents. Using the $Bi_2Te_3$ quintuple layer as an exemplary material, we show that magnetic proximity from $MnBi_2Te_4$ septuple layers yield finite and tunable shift and injection photocurrents. We apply group analysis and first-principles calculations to evaluate the layer-specific shift and injection current generations under linearly polarized light irradiation. We find that the magnetic injection photoconductivity that localized on the $Bi_2Te_3$ layer can reach over $70\times10^8$ A/(V$^2$·s), so that a 1D linear current density on the order of 0.1 mA/nm can be achieved under an intermediate intensity light. In addition to charge current, we also extend our discussions into spin BPV current, giving pure photo-generated spin current. The vertical propagation direction between the charge and spin photocurrents suggest that they can be used individually in a single material. Compared with previously reported methods, the magnetic proximity effect via van der Waals interface does not significantly alter the intrinsic feature of the centrosymmetric material (e.g., $Bi_2Te_3$), and its manipulation can be easily achieved by the proximate magnetic configurations (of $MnBi_2Te_4$), interlayer distance, and light polarization.

---

[*] Email: jianzhou@xjtu.edu.cn

**Introduction.**

Bulk photovoltaic (BPV) effect [1-3] that converts light irradiation into electric current in a single material has been attracting tremendous interests recently, which could avoid the traditional $p-n$ semiconductor heterojunction fabrications [4,5]. In centrosymmetric ($\mathcal{P}$) broken systems, light induces anharmonic motion of carriers (electrons in the conduction bands and holes in the valence bands). Hence, two typical second order nonlinear optical processes, namely, sum frequency and difference frequency generations occur simultaneously [6,7]. The BPV effect arises from the difference frequency generation process under monochromatic irradiation (with angular frequency of $\omega$). Two major mechanisms of BPV processes that have been extensively studied are shift current and injection current generations, both of which originate from the geometric phases of electronic states (such as Berry phase) that strongly rely on symmetry constraints [8-10]. Under time-reversal symmetry $\mathcal{T}$, shift current emerges under linearly polarized light (LPL) illumination,

$$j^a_{\text{NSC}} = \sigma^{a,bc}_{\text{NSC}}(0,\omega,-\omega)E^b(\omega)E^c(-\omega), \qquad (1)$$

where $j^a_{\text{NSC}}$ indicates normal shift current (NSC) density that is induced by light alternating electric field $\mathbf{E}(\omega)$. $\sigma^{a,bc}_{\text{NSC}}(0,\omega,-\omega)$ is the NSC photoconductivity coefficient, with $a$, $b$, and $c$ referring to Cartesian axes. The $\sigma^{a,bc}_{\text{NSC}}(0,\omega,-\omega)$ scales with the shift vector between the valence and conduction bands. On the other hand, the circularly polarized light produces normal injection current (NIC),

$$\frac{dj^a_{\text{NIC}}}{dt} = \eta^{a,bc}_{\text{NIC}}(0,\omega,-\omega)E^b(\omega)E^c(-\omega). \qquad (2)$$

The injection current density grows with time $t$ and saturates at the carrier relaxation time $\tau$. Therefore, one usually discusses the photoconductivity $\eta^{a,bc}_{\text{NIC}}(0,\omega,-\omega)$ for the time derivative of current density $j^a_{\text{NIC}}$. The NIC photoconductivity scales with interband Berry curvature. Hence, both shift and injection currents arise from the quantum topological nature of electronic wavefunctions.

Very recently, it has been disclosed that for magnetic $\mathcal{T}$-broken systems, there are two additional cousin photocurrent mechanisms, namely, magnetic shift current (MSC) and magnetic injection current (MIC). For example, Wang and Qian have shown that these magnetic version photocurrents emerge in antiferromagnetic (AF) MnBi$_2$Te$_4$ bilayers with $\mathcal{PT}$-symmetry [11]. Similar BPV currents have also been evaluated in the AF CrI$_3$ bilayers [12,13] and their fundamental mechanisms have been revealed [14]. In these cases, the symmetry arguments and numerical calculations demonstrate that both NSC and NIC would vanish. For the $\mathcal{PT}$-systems, generally speaking the linearly and circularly polarized light could induce MIC and MSC, respectively. Depending on specific materials, their existence can be determined by group theory (see Ref. [11-14] and Supplemental Material [15], SM for detail discussions).

In order to conceive finite photocurrents, as stated previously, one has to resort to $\mathcal{P}$-broken systems. This excludes a large number of spatially centrosymmetric materials, unless they are chemically doped (or alloyed) [16], structurally deformed [17-20], or electrically biased [21,22]. These schemes could effectively break $\mathcal{P}$ in either transient or permanent strategies, but usually requiring direct contacts with the samples in chemical, electrochemical, and/or mechanical techniques. In this work, we propose that magnetic proximity effect [23] is another facile and controllable method to manipulate the centrosymmetry in van der Waals (vdW) layers and generate large layer-resolved BPV currents. Note that the vdW interactions are sufficiently weak (on the order of 1−10 μJ/cm$^2$) [24,25] and are less susceptible to lattice distortions or damages. The magnetic proximity effect through vdW gap is even marginal (on the order of 0.1 μJ/cm$^2$). We use MnBi$_2$Te$_4$ (denoted as MBT) bilayer and Bi$_2$Te$_3$ (denoted as BT) to build superlattice thin films to illustrate our theory and apply first-principles calculations to evaluate their layer-resolved BPV currents. These superlattices possess the same magnetic point group with bilayer MBT ($\bar{3}'m'$ for interlayer AF and $\bar{3}m'$ for interlayer FM). They have been recently fabricated and host versatile intrinsic magnetic topological features, such as quantum anomalous Hall effect and axion insulating phase [26-31]. The MBT septuple layer (SL) composes seven atomic layers (Te-Bi-Te-Mn-Te-Bi-Te), which are vdW separated by BT qutintuple layers (QLs, Te-Bi-Te-Bi-Te) with small interlayer binding energy of 52 μJ/cm$^2$ (comparable to other transition metal chalcogenides and halides) [32,33]. The Mn 3d magnetic moments favor a long range interlayer AF coupling and a normal intralayer ferromagnetic (FM) magnetic configuration below its Neel temperature (~20 K depending on superlattice structures) [34]. This breaks $\mathcal{T}$ while preserves $\mathcal{PT}$ in the whole system. Even though the BT QLs are crystalline centrosymmetric and intrinsically nonmagnetic, we perform first-principles density functional theory (DFT) calculations and show that sizable and large BPV currents emerge in the BT QL. This is due to the magnetic proximity effect that arises from MBT SLs through the vdW gap, so that the $\mathcal{P}$ is broken in the magnetic point group framework. To be specific, we calculate the NSC and MIC projected onto different vdW layers. At the AF ground state, finite photocurrent emerges in the BT QLs and its MIC photoconductivity can be as large as 72.4×10$^8$ A/(V$^2$·s). The photocurrent density increases with the quadratic power of electric field (or linear relation with light intensity). For example, under a LPL with intensity of 3.3×10$^{10}$ W/cm$^2$ (electric field magnitude of 0.5 V/nm), the photocurrent density reaches 361.8 μA/nm$^2$, assuming a conservative carrier lifetime of 0.2 ps. We also show that the magnetic configuration of MBT SLs controls the magnitude and direction of different photocurrents. The calculations reveal that time-reversal of the AF pattern would reverse the MIC while keeping the NSC, as the former is $\mathcal{T}$-odd and the latter is $\mathcal{T}$-even. If the interlayer magnetic configurations between MBT SLs are driven to be FM, $\mathcal{P}$ preserves rather than $\mathcal{PT}$. The layer-resolved photocurrents in the BT QLs then vanish completely. The $z$-magnetic moment confines the system with $C_{3z}$, hence different LPL polarization direction changes the BPV current as well. We also extend our calculations and discuss the BPV spin current. The spin current has a surplus spin-$S_z$ that transforms as a pseudovector. Under LPL irradiation, it also composes MIC and NSC nature. The former one is $\mathcal{T}$-even, while the latter is $\mathcal{T}$-odd. Therefore, the spin current, in the same manner, can be fine-tuned via both magnetic alignments and light polarization.

**Methods.**

*Density Functional Theory.* The first-principles calculations are implemented in the Vienna *ab initio* simulation package (VASP) [35,36] which adopts projector augmented-wave (PAW) method [37] and planewave basis set (cutoff energy of 350 eV) to treat the core and valence electrons, respectively. The exchange-correlation functional is using the generalized gradient approximation (GGA) in the solid state Perdew-Burke-Ernzerhof (PBEsol) form [38]. The periodic boundary condition is used and the first Brillouin zone is represented by a Γ-centered **k**-mesh with (15×15×1) grid[39]. In order to incorporate the strong correlations on magnetic Mn 3d orbitals, we add an effective Hubbard $U$ (= 5.34 eV) correction according to the Dudarev scheme [40,41]. This value is widely used in previous works and has been demonstrated to yield results consistent with experimental observations [42-44]. One notes that other $U$ values could give qualitatively same results. A vacuum space of at least 15 Å is added in the out-of-plane $z$ direction to eliminate the artificial interaction between different images under periodic boundary condition. Self-consistent spin-orbit coupling interactions have been used in all calculations. The criteria of total energy and force component are set to be $1\times10^{-7}$ eV and 0.01 eV/Å, respectively. The vdW interactions are semi-empirically included in the zero damping DFT-D3 method [45].

*Bulk photovoltaic conductivities.* The BPV photoconductivity coefficients are calculated using the Wannier representation, which includes Mn−d, Bi−p, and Te−p (covering all bands from −6.2 to 4.8 eV relative to the Fermi level) to fit the DFT electronic structures [46-48]. A *k*-mesh grid of 601×601×1 is adopted to integrate the optical conductivities, which is tested to achieve sufficient convergence accuracy. According to the quadratic order Kubo perturbation theory, the BPV photoconductivity can be evaluated in various gauge schemes. The velocity gauge approach, which tends to diverge at low incident frequency, can be simplified into different photocurrents at the long carrier lifetime limit [49-52]. In detail, one can apply band theory to evaluate the NSC and MIC conductance coefficients separately,

$$\sigma_{\text{NSC}}^{a,bc}(0,\omega,-\omega) = \frac{\pi e^3}{2\hbar^2} \int \frac{d^3\mathbf{k}}{(2\pi)^3} \sum_{n,m} f_{nm} \text{Im}\left(r_{mn}^b r_{nm}^{c;a} + r_{mn}^c r_{nm}^{b;a}\right) \delta(\omega_{nm} - \omega) \quad (3)$$

and

$$\eta_{\text{MIC}}^{a,bc}(0,\omega,-\omega) = -\frac{\pi e^3}{2\hbar^2} \int \frac{d^3\mathbf{k}}{(2\pi)^3} \sum_{n,m} f_{nm} \Delta_{nm}^a \left(r_{mn}^b r_{nm}^c + r_{mn}^c r_{nm}^b\right) \delta(\omega_{mn} - \omega) \quad (4)$$

Here the position operator matrix is defined as $r_{nm}^a = \frac{v_{nm}^a}{i\omega_{nm}} = \frac{\langle n|\frac{\partial H}{\partial k^a}|m\rangle}{i\omega_{nm}}$ ($n \neq m$) with $\omega_{nm} = \omega_{nn} - \omega_{mm}$ measuring the band energy difference between band $n$ and $m$. $f_{nm} = f_n - f_m$ and $\Delta_{nm}^a = v_{nn}^a - v_{mm}^a$ are occupation and velocity difference, respectively. All the quantities that depend on reciprocal space coordinate **k**, and such **k**-dependence is omitted in these equations for clarity reason. The Dirac delta function is represented by Lorentz function with a broadening factor of 0.02 eV, which guarantees the energy conservation law. In order to analyze the integrand of NSC and MIC, we denote them as $\varsigma(\mathbf{k},\omega) = \sum_{n,m} f_{nm} \text{Im}\left(r_{mn}^b r_{nm}^{c;a} + r_{mn}^c r_{nm}^{b;a}\right) \delta(\omega_{nm} - \omega)$ and $\zeta(\mathbf{k},\omega) = \sum_{n,m} f_{nm} \Delta_{nm}^a \left(r_{mn}^b r_{nm}^c + r_{mn}^c r_{nm}^b\right) \delta(\omega_{mn} - \omega)$, respectively. The sum rule for derivative of **r** is

$$r_{nm}^{b;a} = \frac{\partial r_{nm}^b}{\partial k^a} - i(\mathcal{A}_{nn}^a - \mathcal{A}_{mm}^a) r_{nm}^b$$

$$= \frac{i}{\omega_{nm}} \left[ \frac{v_{nm}^b \Delta_{nm}^a + v_{nm}^a \Delta_{nm}^b}{\omega_{nm}} - w_{nm}^{ba} + \sum_{l \neq n,m} \left( \frac{v_{nl}^b v_{lm}^a}{\omega_{lm}} - \frac{v_{nl}^a v_{lm}^b}{\omega_{nl}} \right) \right], \quad (5)$$

where $w_{nm}^{ba} = \langle n|\frac{\partial^2 H}{\partial k^b \partial k^a}|m\rangle$. Both of $\sigma_{\text{NSC}}^{a,bc}(0,\omega,-\omega)$ and $\eta_{\text{MIC}}^{a,bc}(0,\omega,-\omega)$ feature electronic topological characters. One defines the shift vector under LPL irradiation between two bands as

$$R_{nm}^{a,b} = \frac{\partial \phi_{nm}^b}{\partial k^a} + \mathcal{A}_{nn}^a - \mathcal{A}_{mm}^a, \quad (6)$$

where $\phi_{nm}^b$ is the complex phase of $r_{nm}^b$ ($= |r_{nm}^b| e^{-i\phi_{nm}^b}$). Therefore, it can be shown that $R_{nm}^{a,b} |r_{nm}^b|^2 = \text{Im}(r_{mn}^b r_{nm}^{b;a})$ which transforms as a polar vector. On the other hand, the quantum metric tensor is defined as

$$g_{nm}^{bc} = \{r_{mn}^b, r_{nm}^c\} = (r_{mn}^b r_{nm}^c + r_{mn}^c r_{nm}^b), \quad (7)$$

which demonstrates the topological feature of MIC. Note that for $\mathcal{T}$-symmetric systems, the MIC vanishes (under LPL), while the NSC is symmetrically forbidden in $\mathcal{PT}$-systems.

As for spin photocurrent, we adopt the general definition of $\hat{\mathfrak{V}}^{ad} = \frac{\{\hat{S}_d, \hat{v}^a\}}{2} = \frac{1}{2}(\hat{S}_d \hat{v}^a + \hat{v}^a \hat{S}_d)$, where $\hat{S}_d$ is the spin operator ($d$ is taken to be $z$ in the current work). Hence, in order to evaluate the spin current, we replace the velocity operator $\hat{v}^a$ by $\hat{\mathfrak{V}}^{ad}$. For example, the spin-velocity derivative can be written as [53]

$$d_{nm}^{bd;a} = \frac{1}{2}\langle n|\{S_d, w^{ab}\}|m\rangle + \sum_{p\neq n}\frac{\mathfrak{V}_{np}^{ad} v_{pm}^b}{\omega_{np}} + \sum_{p\neq m}\frac{v_{np}^b \mathfrak{V}_{pm}^{ad}}{\omega_{mp}}. \quad (8)$$

It would reduce to the charge current formula when one takes the spin operator $S_d$ into an identify matrix form. Since the MIC only incorporates the intraband contribution of spin current, we can include spin torque effect [19]. This is consistent with the "proper" definition of spin current via $\hat{\mathfrak{V}}^{ad} = \frac{d(\hat{S}_d \hat{r}^a)}{dt}$ [54]. In this case, the Onsager relation is maintained, so that the spin current obeys the conservation law of equation of motion.

In order to evaluate the layer-resolved current, one can introduce a layer projector operator, $\hat{P}_L = \sum_{i \in L} |\phi_i\rangle\langle\phi_i|$ with Wannier function $|\phi_i\rangle$ localized on a specific QL or SL [55]. Then we multiple this $\hat{P}_L$ with velocity (or spin current) operator $\hat{v}^a$ (or $\hat{\mathfrak{V}}^{ad}$), through which one could map the photocurrent onto each vdW layer. We have checked that the summation of each layer-resolved current on all the layers are almost the same as the total photocurrent. Hence, the interlayer photocurrent is marginal, as the vdW gap is ~3 Å. Note that in Eqs. (3) and (4), all the **k**-space integrals are taken in the 3D Brillouin zone, and the vacuum contributions have been included. We thus rescale these in-plane photoconductivity coefficients $\sigma_{\text{NSC}}^{a,bc}(0,\omega,-\omega)$ and $\eta_{\text{MIC}}^{a,bc}(0,\omega,-\omega)$ by multiplying a factor $\frac{L_z}{d_{\text{eff}}}$, where $L_z$ is the simulation supercell lattice constant along $z$ and $d_{\text{eff}}$ is an effective thickness of the vdW layer [56,57]. In the current work, we use $d_{\text{eff}}$ of MBT SL and BT QL to be 14 and 11 Å, respectively, which are taken from the thickness values when they form vdW bulk structures.

**Results.**

*Photocurrents of MBT-MBT bilayer ($Mn_2Bi_4Te_8$).* We first examine the atomic geometries, electronic behaviors, and magnetic configurations of different MBT and BT superlattice slabs. In order to maintain the completeness of our study, we first explore the MBT bilayer (MBT-MBT) without BT intercalation [Fig. 1(a)]. The Te atoms serve as termination layers between the MBT, leaving a vdW gap of 2.9 Å. Each Mn carries ~5 $\mu_B$ (Bohr magneton) magnetic moments pointing along the easy axis $z$, and the intralayer coupling is strongly locked to be FM. The interlayer MBT magnetic coupling prefers an AF configuration. This is usually terminated as the A-type AF state. We reveal that the magnetic interlayer exchange energy $\Delta E = E_{FM} - E_{AF}$ ($E$ refers to total energy) is 0.43 meV per unit cell (0.04 $\mu J/cm^2$). According to previous experimental facts, this AF ground state can be converted into FM via applying an external magnetic field below its Neel temperature (~12 T at 4.5 K or 6 T at 1.6 K) [58].

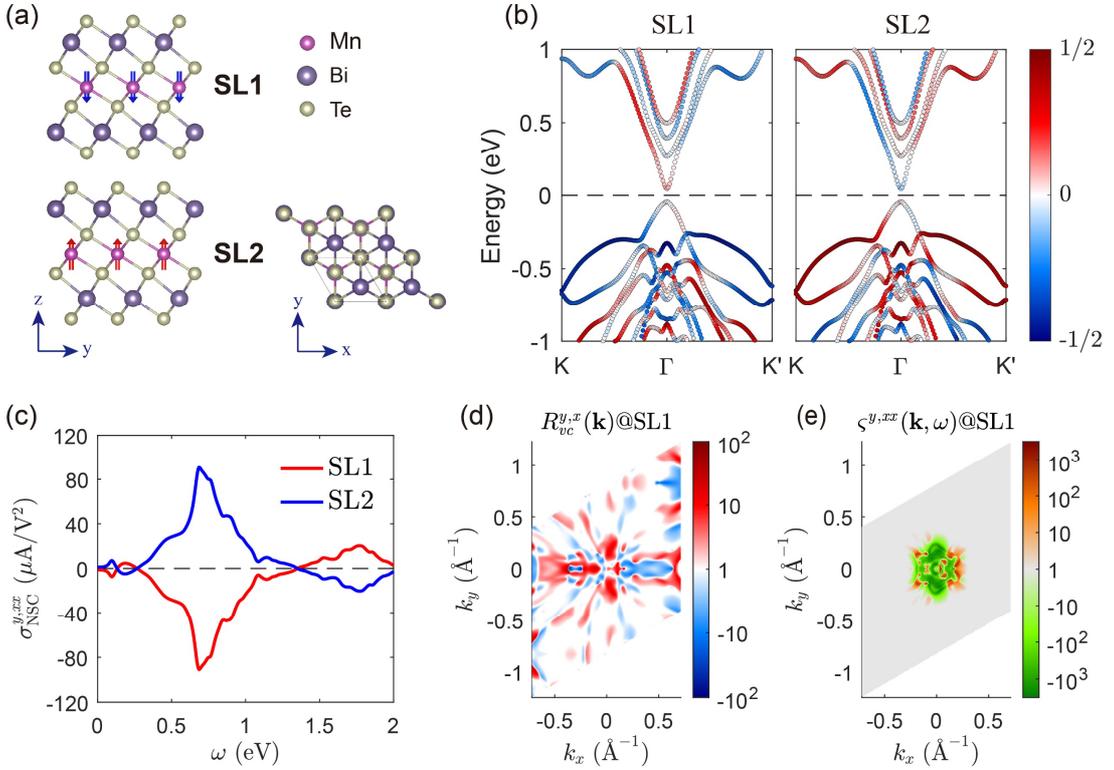

FIG. 1. Structure and normal shift current of the MBT-MBT bilayer. (a) Atomic geometric structure of a MBT-MBT bilayer, with arrows in the central Mn atoms representing (AF) magnetic moments. The SL1 and SL2 are denoted. Both sideview

and top view structures are shown. (b) Band dispersion along the high symmetric **k**-path, where different color represents spin polarization projected onto the two SLs, the spin-orbit coupling effect is considered. (c) The calculated NSC on different MBT SLs under *x*-polarized LPL. The abscissa axis denotes the incident photon frequency ($\hbar$ = 1). Only NSC flowing along *y* is allowed, while the *x*-current is symmetrically forbidden (not shown). (d) Shift vector $R_{vc}^{y,x}(\mathbf{k})$@SL1 distribution over the first BZ between the top valence ($v$) and bottom conduction ($c$) band. (e) **k**-resolved NSC integrand $\varsigma^{y,xx}(\mathbf{k},\omega)$@SL1 at the incident energy of $\omega = 0.69$ eV where NSC presents a peak.

Figure 1(b) depicts the calculated electronic band dispersion along the high symmetric **k**-path, in the AF state. One sees a semiconducting feature with a direct bandgap of 88 meV, where both valence band maximum (VBM) and conduction band minimum (CBM) locate at the Γ point. The band structure without SOC effect is shown in Figure S1 [15], which gives a larger direct bandgap. This is consistent with the results in previous works [59]. In the AF magnetic configuration ($\mathcal{PT}$), each electronic band state is doubly degenerate without net spin polarization. The layer-dependent spin polarization exhibits a hidden feature. In our results, we use color scheme to illustrate spin polarizations $\langle S_{n\mathbf{k}}^z \rangle$ locally contributed from the SL1 and SL2, which shows opposite values in the two vdW SLs. Note that in our calculations, we have summed the doubly degenerated state contributions for each $n$ at **k**.

We next focus on the NSC and MIC photoconductivities. As discussed previously [60], symmetry operators pose constraints on both photocurrent direction and magnitude. Here we perform similar analysis on the two MBT SLs separately. The coordinates of each MBT SL contain a mirror reflection that is normal to *x*-direction, $\mathcal{M}_x$ [Fig. 1(a)]. Since Mn is magnetic along *z*, one has to multiple an additional time-reversal operator, which gives $\mathcal{M}_x\mathcal{T}$. Thus, the local group on one MBT SL is reduced to be $3m'$, breaking the $\mathcal{PT}$-symmetry. It then yields sizable NSC generation with one independent component (see SM)[15]. In detail, the $\mathcal{M}_x\mathcal{T}$ maps BZ coordinate $(k_x, k_y)$ to $(k_x, -k_y)$, which also constrains the shift vector via $\mathcal{M}_x\mathcal{T} R_{nm}^{x,b}(k_x, k_y) = -R_{nm}^{x,b}(k_x, -k_y)$, where $b = x, y$. Hence, under *x*-LPL (or *y*-LPL) irradiation, the NSC is symmetrically forbidden along *x*, leaving $\sigma_{\text{NSC}}^{y,xx}(0,\omega,-\omega) \neq 0$. On the other hand, the MIC $\eta_{\text{MIC}}^{a,bb}(0,\omega,-\omega)$ is determined by velocity difference $\Delta_{nm}^a$ and quantum metric $g_{nm}^{bb}$ between bands $n$ and $m$. Under $\mathcal{M}_x\mathcal{T}$, one has $\Delta_{nm}^y(k_x,k_y) = -\Delta_{nm}^y(k_x,-k_y)$ and $g_{nm}^{bb}(k_x,k_y) = g_{nm}^{bb}(k_x,-k_y)$. In this regard, MIC flows along *x* and must vanish along *y*, namely, $\eta_{\text{MIC}}^{y,bb}(0,\omega,-\omega) = 0$. These discussions are consistent with magnetic group theory results. Therefore, we suggest that the current directions of NSC and MIC under *x*-LPL would be vertical, so that their detection and measurement can be performed individually. We tabulate the spatial and time-reversal transformation of various gauge-invariant quantities in Table I.

Table I. Transformation symmetries of different quantities (**k** coordinate, shift vector, quantum metric, velocity difference, and spin operator) under spatial and time-reversal operators.

|  | $\mathcal{M}_i$ | $\mathcal{P}$ | $\mathcal{T}$ | $\mathcal{M}_i\mathcal{T}$ | $\mathcal{PT}$ |
|---|---|---|---|---|---|
| $(k_x,k_y)$ | $(-k_x,k_y)$ | $(-k_x,-k_y)$ | $(-k_x,-k_y)$ | $(k_x,-k_y)$ | $(k_x,k_y)$ |
| $R_{nm}^{a,b}$ | $(-1)^{\delta_{ia}}R_{nm}^{a,b}$ | $-R_{nm}^{a,b}$ | $R_{nm}^{a,b}$ | $(-1)^{\delta_{ia}}R_{nm}^{a,b}$ | $-R_{nm}^{a,b}$ |
| $g_{nm}^{bc}$ | $(-1)^{\delta_{ib}+\delta_{ic}}g_{nm}^{bc}$ | $g_{nm}^{bc}$ | $g_{nm}^{bc}$ | $(-1)^{\delta_{ib}+\delta_{ic}}g_{nm}^{bc}$ | $g_{nm}^{bc}$ |
| $\Delta_{nm}^a$ | $(-1)^{\delta_{ia}}\Delta_{nm}^a$ | $-\Delta_{nm}^a$ | $-\Delta_{nm}^a$ | $(-1)^{\delta_{ia}+1}\Delta_{nm}^a$ | $\Delta_{nm}^a$ |
| $S_{nm}^d$ | $(-1)^{\delta_{id}+1}S_{nm}^d$ | $S_{nm}^d$ | $-S_{mn}^d$ | $(-1)^{\delta_{id}}S_{mn}^d$ | $-S_{mn}^d$ |

We also note that the two MBT SLs can be mapped via $\mathcal{PT}\mathbf{x}_{\text{SL1}} = \mathbf{x}_{\text{SL2}}$ (**x** includes both spatial coordinates and local magnetic moments). Since the NSC is $\mathcal{PT}$-odd, the NSC localized in the two MBT SLs are opposite, namely, $\sigma_{\text{NSC}}^{y,xx}(0,\omega,-\omega)@\text{SL1} = -\sigma_{\text{NSC}}^{y,xx}(0,\omega,-\omega)@\text{SL2}$. This is consistent with the fact that NSC vanishes in the whole AF MBT bilayer system. The MIC generation, on the other hand, is $\mathcal{PT}$-even. Then, we have $\eta_{\text{MIC}}^{x,xx}(0,\omega,-\omega)@\text{SL1} = \eta_{\text{MIC}}^{x,xx}(0,\omega,-\omega)@\text{SL2}$.

In order to verify these symmetry analyses, we plot our first-principles calculation results in Figs. 1(c)−1(e). Consistent with our argument, the $\sigma_{NSC}^{x,xx}(0,\omega,-\omega)$ is always zero, which is not shown for clarity reason. The $\sigma_{NSC}^{y,xx}(0,\omega,-\omega)$ on the SL1 reaches −90.9 μA/V² at an incident photon energy $\omega$ = 0.69 eV [Fig. 1(c)]. Note that this value surpasses the NSC photoconductivity calculated in many other 2D materials, such as MoS$_2$ [61], GeS [60], and α-NP [62]. Hence, an x-LPL light at this photon energy with alternating electric field magnitude $E$ = 0.5 V/nm could trigger a current density of −22.7 μA/nm²; the negative sign indicating current along −y. In order to clearly show the BZ contribution, we depict the shift vector $R_{vc}^{y,x}(\mathbf{k})@SL1$, where clear symmetries between $(k_x, k_y)$ and $(k_x, -k_y)$ can be observed [Fig. 1(d)], confirming its transformation as a polar vector. The NSC integrand $\varsigma^{y,xx}(\mathbf{k},\omega)$ (see Methods) is plotted [Fig. 1(e)], which suggests that the NSC is mainly contributed from the vicinity of the Γ point. The other SL contributes oppositely to the NSC (with an opposite shift vector). As a result, the net NSC of MBT-MBT bilayer is zero, which has been largely overlooked previously. Nevertheless, in this work, we propose that there emerges hidden layer-dependent NSC current with a large magnitude. The vdW gap that separates the two SLs allows one to detect such NSC current with well-deposited electrodes.

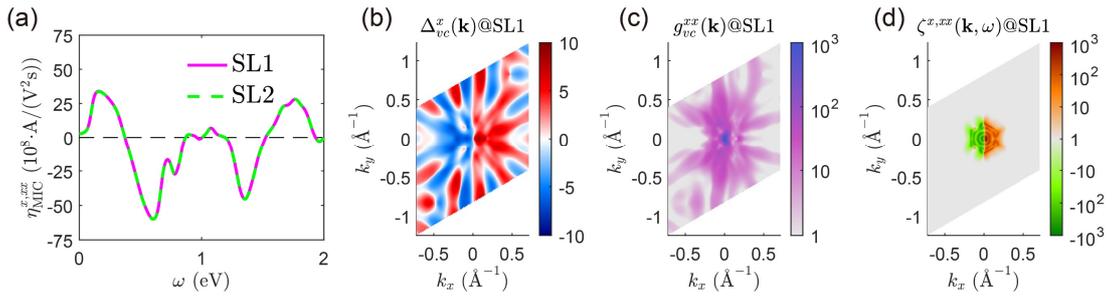

FIG. 2. Magnetic injection current in the MBT-MBT bilayer. (a) The calculated MIC on MBT SLs under x-polarized LPL. Only the x-flowing current is finite, while the y-current is forbidden. (b) Velocity difference $\Delta_{vc}^x(\mathbf{k})@SL1$ and (c) quantum metric distribution $g_{vc}^{xx}(\mathbf{k})@SL1$ over the first BZ. (d) $\mathbf{k}$-resolved MIC integrand $\zeta(\mathbf{k},\omega)@SL1$ at $\omega$ = 0.61 eV. In these plots, the symmetry of $(k_x, k_y)$ and $(k_x, -k_y)$ are clearly seen.

As for the MIC generation, $\eta_{\text{MIC}}^{x,xx}(0,\omega,-\omega)$ on both the two SLs contributed equally [Fig. 2(a)], while the calculated $\eta_{\text{MIC}}^{y,xx}(0,\omega,-\omega)$ is always zero. The calculated $\eta_{\text{MIC}}^{x,xx}(0,\omega,-\omega)$ are consistent with previous works, further verifying our numerical procedure. Adding two SL values yields a total MIC of $-60.0\times10^8$ A/(V$^2$·s) (at an incident photon energy of $\omega = 0.61$ eV). One could then expect a current density of $-300.0$ μA/nm$^2$ under $x$-LPL with $E = 0.5$ V/nm (at the same photon energy), assuming an effective electron lifetime of 0.2 ps. Note that this value is conservative and could be even longer, depending on the sample quality and environmental condition. The **k**-resolved velocity difference between the top valence and bottom conduction bands are plotted [Fig. 2(b)]. Symmetry argument of $\Delta_{vc}^x(k_x,k_y) = \Delta_{vc}^x(k_x,-k_y)$ is clearly observed. Note that the two SLs possess same $\Delta_{vc}^x(\mathbf{k})$. The quantum metric shows similar pattern [Fig. 2(c)]. The integrand of MIC, $\zeta(\mathbf{k},\omega)$@SL1, is plotted for $\omega = 0.61$ eV, where large contributions in the vicinity of Γ is seen [Fig. 2(d)].

In addition to charge current that uses electric charge as the degree of freedom to carry information, electronic spin can serve as another information carrier, which is facile to tune and easy to detect as well. In some recent works, photon illumination generated spin current has been predicted and evaluated in various material platforms [60,61]. Here we extend the LPL-induced charge photocurrent into spin photocurrent that remains further discussion or evaluation in $\mathcal{PT}$-systems [63]. We adopt the anticommutation definition of spin current operator that has been widely used during the past decade. One has to note that spin torque effect may arise when spin operator is not a good quantum number under SOC [54,64]. This could be small in magnetic materials for intrinsic spin polarization. Under LPL, the spin current also composes MIC and NSC nature, which are depicted in Figure 3. One sees that the spin NSCs are the same for both SLs and flow along $y$, $\sigma_{\text{NSC}}^{y_z,xx}(0,\omega,-\omega)$@SL1 = $\sigma_{\text{NSC}}^{y_z,xx}(0,\omega,-\omega)$@SL2 $\neq 0$, while the $\sigma_{\text{NSC}}^{x_z,xx}(0,\omega,-\omega) = 0$ under $\mathcal{M}_x\mathcal{T}$ symmetry constraints. This is because spin operator $\hat{S}_z$ is symmetrically unchanged under $\mathcal{M}_x\mathcal{T}$. Similarly, the spin MICs are opposite in the two SLs, which flow along $x$, namely,

$\eta_{\text{MIC}}^{x_z,xx}(0,\omega,-\omega)@\text{SL1} = -\sigma_{\text{MIC}}^{x_z,xx}(0,\omega,-\omega)@\text{SL2} \neq 0$, yielding zero net spin MIC in the whole system. Therefore, similar as charge current, the spin NSC and MIC are also flowing vertically to each other. When one sums the two SLs together, the charge MIC survives which transports along $x$, while the spin NSC is nonzero, flowing along $y$. This arises pure spin current, that carries no charge carriers and has been playing an essential role for the low energy cost and high efficiency spintronics [65,66].

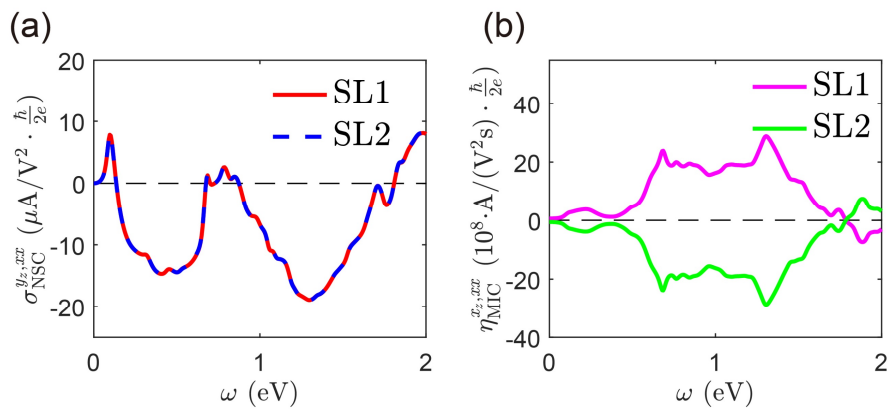

FIG. 3. Spin photocurrents in the MBT-MBT bilayer. (a) The spin NSC photo-conductance in the two SLs under $x$-LPL. (b) The spin MIC photo-conductance in the two SLs under $x$-LPL. The spin NSC flowing along $x$ ($\sigma_{\text{NSC}}^{x_z,xx}$) and spin MIC flowing along $y$ ($\eta_{\text{MIC}}^{y_z,xx}$) are both symmetrically forbidden.

We next show that magnetic configuration could tune these photocurrents. If the time-reversal AF configuration is used, the charge NSC remains their values, since they are $\mathcal{T}$-even. At the same time, the charge MIC reverses its direction, being a $\mathcal{T}$-odd quantity. However, for the spin current, since $S_z$ adds another $\mathcal{T}$-odd symmetry, the spin NSC remains its value while spin MIC becomes opposite. We plot our calculations in Fig. S2 [15]. The FM interlayer configuration, which has been successfully achieved in recent experiments, is equivalent to posing time-reversal $\mathcal{T}$ onto one of these two SLs. Generally speaking, it reverses the charge MIC (and spin NSC) in such SL, while keeping charge NSC (and spin MIC). Note that the exact photoconductivity values are slightly changed due to band structure is altered (Fig. S3) [15]. In this regard, all the charge and spin currents become opposite in the two SLs, giving zero net (charge and spin) NSC and MIC. This is consistent with the fact that the FM MBT-MBT bilayer is $\mathcal{P}$-symmetric. These results are also plotted in Fig. S4 [15]. Therefore, we propose that one could apply magnetic field to toggle and achieve various layer-resolved charge and spin photocurrents.

*Photocurrents of MBT-BT-MBT trilayer ($Mn_2Bi_6Te_{11}$)*. We then intercalate a BT QL inside the MBT-MBT bilayers, forming a MBT-BT-MBT trilayer [Fig. 4(a)]. The two MBT SLs prefer a weak AF configuration, which is energetically lower than the interlayer FM state by 0.12 meV per unit cell. The significantly reduced magnetic exchange energy is ascribed by the long-range coupling between the two MBT SLs. The band dispersion is plotted in Fig. 4(b), where contributions from the BT QL are highlighted by colormap. The direct bandgap becomes 51 meV, and shows no spin polarization feature owing to the $\mathcal{PT}$-symmetry in the whole system. We verify that the local magnetic moments on the BT QL vanish, under the magnetic proximity of two AF aligned MBT SLs.

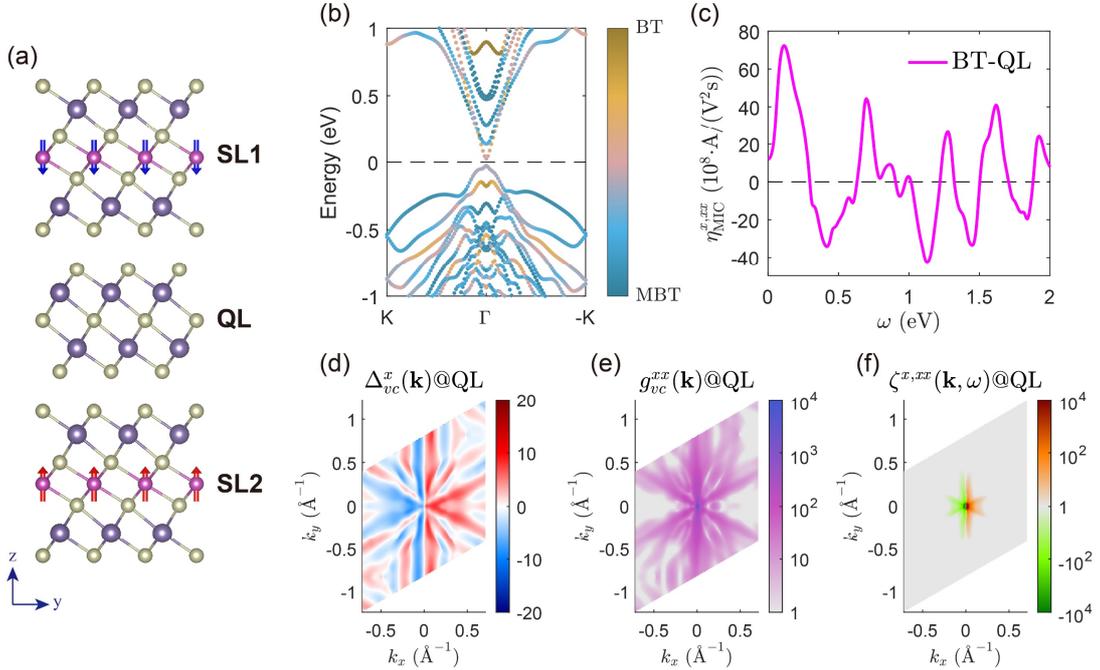

FIG. 4. Photocurrents in MBT-BT-MBT trilayer under *x*-LPL. (a) Geometric structure of MBT-BT-MBT trilayer, with arrows indicating local magnetic moments on the Mn atoms. (b) Band dispersion along the high symmetric **k**-path with contributions from BT QL highlighted. (c) The MIC photoconductivity of the central BT QL. Brillouin

zone distribution of (d) $\Delta_{vc}^{x}(\mathbf{k})$@QL and (e) $g_{vc}^{xx}(\mathbf{k})$@QL between the valence and conduction bands. (f) **k**-resolved $\zeta^{x,xx}(\mathbf{k},\omega)$@QL at $\omega$ = 0.11 eV.

We plot the calculated charge NSC $\sigma_{\text{NSC}}^{y,xx}(0,\omega,-\omega)$ and MIC $\eta_{\text{MIC}}^{x,xx}(0,\omega,-\omega)$ on the two MBT SLs in Fig. S5 [15]. Similar as in the MBT-MBT bilayer case, the former one is opposite in the two MBT SLs, while the latter one shows the same value (and magnitude) on both the SLs. Both of them lie on the same order of magnitude as those in the MBT-MBT bilayer system. Interestingly, we find that the central BT QL exhibits a large MIC type photocurrent [Fig. 4(c)], even though the BT QL is spatially centrosymmetric and show no local magnetizations on the Bi and Te atoms. The magnetic proximity effects from MBT SLs break the spatial symmetry magnetically (magnetic point group of $\bar{3}'m'$), via vdW interfacial coupling. One has to note that unlike intrinsically magnetic systems (such as MBT-MBT bilayer [11], AF CrI$_3$-CrI$_3$ bilayer, and bulk α-Fe$_2$O$_3$ [63]), here the BT QL does *not* show finite local magnetic moments. The calculated MIC reaches 72.4×10$^8$ A/(V$^2$·s) (at incident energy $\omega$ = 0.11 eV) and flows along *x*, which is much larger than that calculated values in the CrI$_3$ bilayer. Thus, a light with 3.3×10$^{10}$ W/cm$^2$ could inject a current density of 361.8 μA/nm$^2$ (assuming carrier lifetime of 0.2 ps). We plot the $\Delta_{vc}^{x}(\mathbf{k})$@QL and $g_{vc}^{xx}(\mathbf{k})$@QL distributions in Figs. 4(d) and 4(e), which yield large contributions around the Γ point. The mirror symmetry is clearly observed. Together with joint density of states, one obtains the $\zeta(\mathbf{k},\omega)$@QL texture [Fig. 4(f)]. Since the intercalated BT QL is $\mathcal{PT}$-symmetric, the NSC would vanish. This is also consistent with our numerical calculations.

In order to further demonstrate the interfacial magnetic proximity effect, we artificially move the two MBT SLs away from the BT QL and calculate photocurrents. The results are plotted in Figure S6 [15]. It shows that as the interlayer distance gradually increases from 2.9 Å (equilibrium) to 6 Å, the magnitude of MIC reduces but with evident values. When the MBT is sufficiently apart from BT QL, the MIC vanishes.

It clearly suggests that the interfacial magnetic proximity is the key mechanism to break $\mathcal{P}$ in this situation, unlike previously reported routes.

We also plot the spin photocurrent of BT QL under *x*-LPL. According to symmetric argument, it has spin NSC nature and flows along *y*. Other spin BPV photocurrents are symmetrically forbidden. The results are shown in Figure 5. The perpendicular flow direction of charge and spin current indicates a pure spin current nature, similar as in other 2D nonmagnetic systems (such as MoS$_2$ [61] and GeS [60]). We see that at $\omega$ = 1.36 eV, the $\sigma_{\text{NSC}}^{y_z,xx}(0,\omega,-\omega)$@QL reaches $-59.7$ μA/V$^2\cdot(\hbar/2e)$, surpassing that calculated values in the CrI$_3$ bilayer [63] and BaFeO$_3$ [67]. The distribution of $\varsigma(\mathbf{k},\omega)$@QL show clear mirror symmetry.

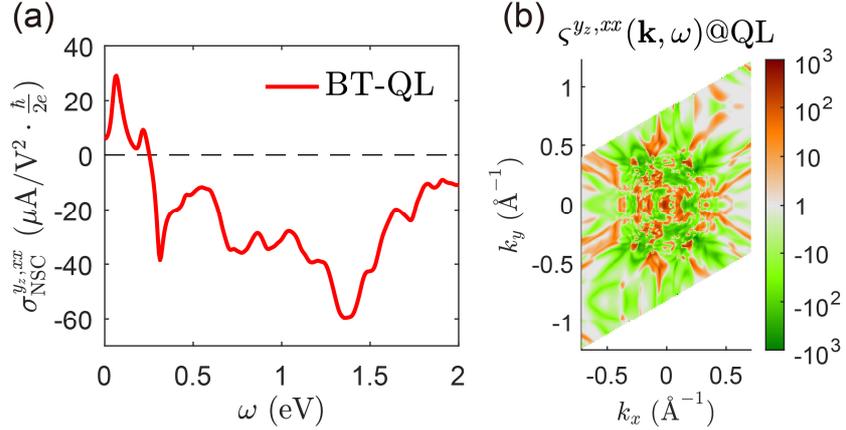

FIG. 5. Spin photocurrents on BT QL in the MBT-BT-MBT trilayer under *x*-LPL. (a) Spin NSC of BT $\sigma_{\text{NSC}}^{y_z,xx}(0,\omega,-\omega)$@QL. (b) *k*-resolved $\varsigma(\mathrm{k},\omega)$@QL at $\omega$ = 1.36 eV.

The magnetic order of MBT SLs also tunes the layer-specific photocurrents. The time-reversal of AF could reverse the charge MIC and spin NSC at BT QL (Fig. S7) [15], as $\mathcal{T}$ flips the sign of $\Delta_{nm}^x$ and $S_{nm}^z$, respectively. Recent experiments show that the MBT SLs can transit into FM alignment under a much weaker magnetic field (0.22 T at 2 K) [26], owing to the long range coupling between them. In the FM state, the symmetry of MBT-BT-MBT becomes $\mathcal{P}$, then only charge and spin photocurrents appear on the MBT SLs, while the BT is completely silent (Fig. S8) [15] even though it is slightly magnetized.

*Photocurrents of MBT-BT-BT-MBT tetralayer ($Mn_2Bi_8Te_{14}$)*. Finally, we propose that such approach can be extended into thicker superlattice thin films, such as MBT-BT-BT-MBT tetralayer [Fig. 6(a)]. We plot the charge and spin photocurrents on the MBT SLs in Fig. S9 [15], and those localized on the two BT QLs (QL1 and QL2) in Figs. 6(b)−6(e). Similarly, the charge NSC along *x* and MIC along *y* are symmetrically forbidden. We see that $\sigma_{\text{NSC}}^{y,xx}(0,\omega,-\omega)@\text{QL1} = -\sigma_{\text{NSC}}^{y,xx}(0,\omega,-\omega)@\text{QL2} \neq 0$ and $\eta_{\text{MIC}}^{x,xx}(0,\omega,-\omega)@\text{QL1} = \eta_{\text{MIC}}^{x,xx}(0,\omega,-\omega)@\text{QL2} \neq 0$. The magnitude of these photocurrents are large enough to be observed experimentally, suggesting the capacity of $\mathcal{P}$-broken under magnetic proximity effect. As for the spin current generation, the spin NSC satesfies $\sigma_{\text{NSC}}^{y_z,xx}(0,\omega,-\omega)@\text{QL1} = \sigma_{\text{NSC}}^{y_z,xx}(0,\omega,-\omega)@\text{QL2} \neq 0$ (the *x*-direction spin NSC is forbidden) and spin MIC is $\eta_{\text{MIC}}^{x_z,xx}(0,\omega,-\omega)@\text{QL1} = -\eta_{\text{MIC}}^{x_z,xx}(0,\omega,-\omega)@\text{QL2} \neq 0$ (the *y*-direction spin MIC is forbidden), which generally is normal to the charge current. If one adds the QL1 and QL2 responses together, it is qualitatively the same as that of BT QL in the MBT-BT-MBT system. Note that different magnetic configurations in the MBT SLs could also control the charge and spin photocurrents localized on specific vdW layers (Fig. S10) [15].

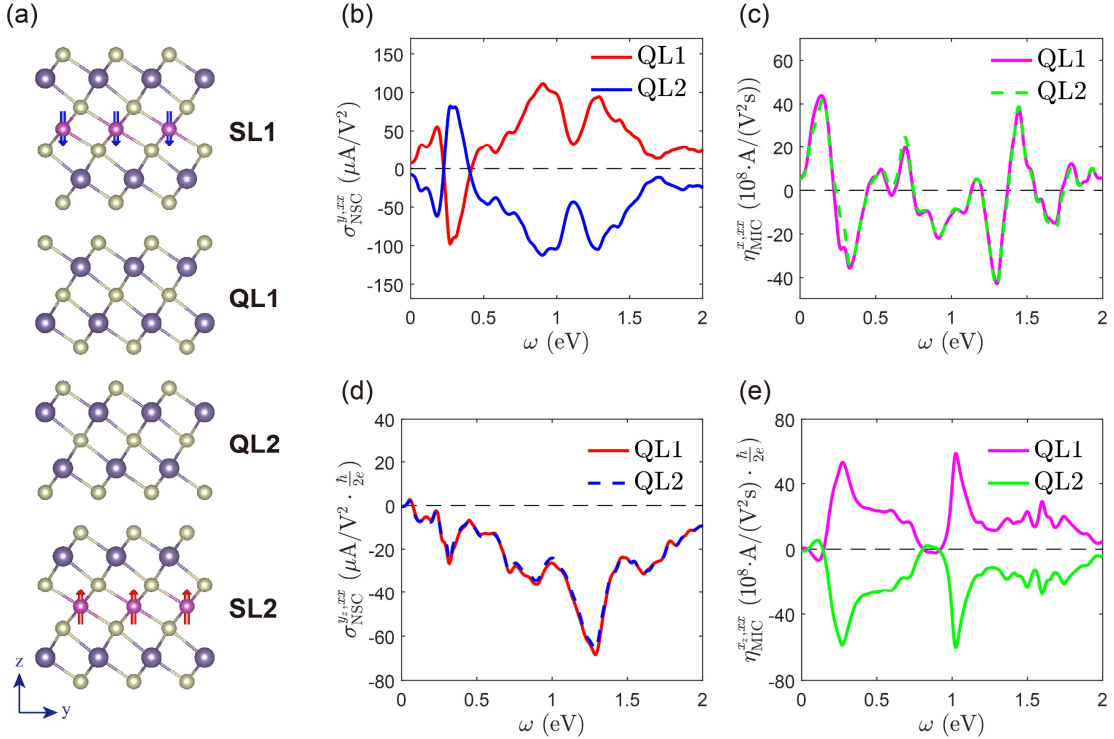

FIG. 6. Charge and spin photocurrents on QLs in the MBT-BT-BT-MBT tetralayer under *x*-LPL. (a) Atomic geometric structure and magnetic alignment of the MBT-BT-BT-MBT tetralayer. (b) Charge NSC of the two BT QLs, $\sigma_{NSC}^{y,xx}(0,\omega,-\omega)$@QL1 and $\sigma_{NSC}^{y,xx}(0,\omega,-\omega)$@QL2, showing opposite flowing directions with same magnitude. (c) Charge MIC $\eta_{MIC}^{x,xx}(0,\omega,-\omega)$@QL1 and $\eta_{MIC}^{x,xx}(0,\omega,-\omega)$@QL2, which are the same in both magnitude and direction. (d) Spin NSC $\sigma_{NSC}^{y_z,xx}(0,\omega,-\omega)$ and (e) spin MIC $\eta_{MIC}^{x_z,xx}(0,\omega,-\omega)$ localized on the two BT QLs.

**Discussion.**

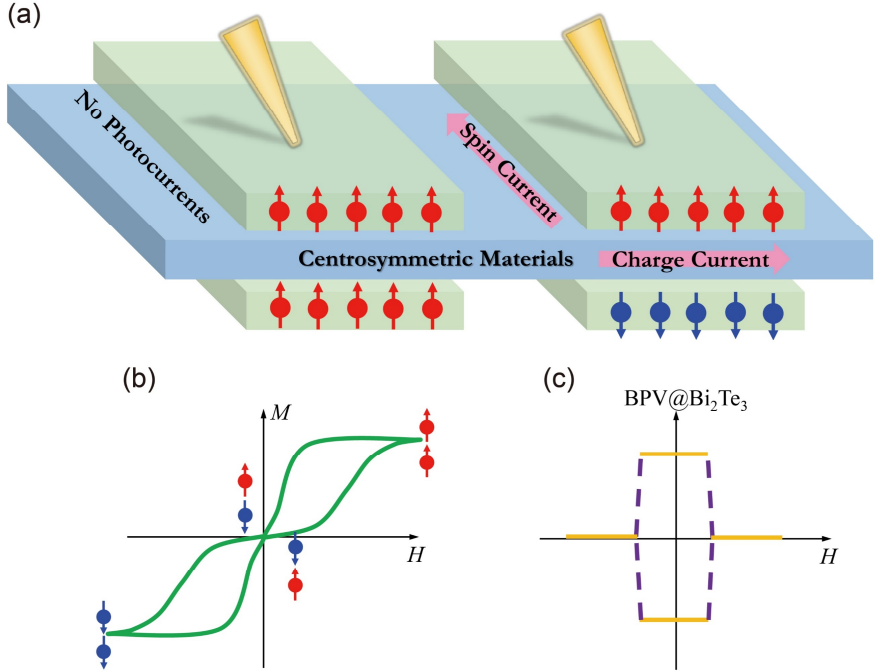

FIG. 7. Schematic plot of BPV currents measurement. (a) AF sandwich could generate BPV currents in spatially centrosymmetric layers, which is silent intrinsically. (b) Magnetic hysteresis loop of AF system, and (c) plots the magnetic field controlled BPV photoconductivity.

We schematically depict the mechanism and propose the potential measurement setup in Fig. 7. As demonstrated previously, the charge and spin currents flow vertically [Fig. 7(a)], which become silent after removing the AF sandwich layers, or when they form FM configuration. One notes that the AF configuration can be fine-tuned via external magnetic field, forming a double hysteresis loop on reversing the field sweep [Fig. 7(b)]. Accordingly, the charge and spin BPV currents would emerge and diminish with respect to proximate magnetic orders [Fig. 7(c)].

The photoconductivity that arises from the magnetic SL also provides a facile way to detect and measure axion insulating phase that usually appears in AFM layered materials. In axion insulators, the top and bottom layers that contribute $1/2$ and $-1/2$ Chern numbers can be connected via a $\mathcal{PT}$ operator. The total Chern number is zero in an intrinsic axion insulator, which would provide a nonzero Chern number when $\mathcal{T}$ is broken under external magnetic field. Here, we show that the charge MIC is even under $\mathcal{PT}$. Hence, the total MIC is nonzero in an intrinsic axion insulator. This would provide an additional route for measuring axion insulators and does not require a magnetic field.

In our above evaluations, we only focused on the *x*-polarized LPL. Since the system is $C_{3z}$ rotational invariant, the *y*-polarized LPL irradiation would induce similar (charge and spin) BPV photocurrents with a sign change. Hence, the carrier motion can be effectively manipulated by the polarization direction of LPL (obeying a sinusoidal function with respect to polarization angle). As for the circularly polarized light, it would introduce additional photocurrent via its phase factor. Generally, such phase factor results in 90°-rotated photocurrents. The shift and injection nature varies as well.

We summarize the mirror reflection, spatial central inversion, and time-reversal impacts on the LPL-induced charge and spin photocurrents in Table II. The "even" and "odd" indicate that under a specific operation, whether such BPV current would keep or flip its sign, respectively. The combination of them can be easily deduced. For example, the $\mathcal{PT}$-symmetry assigns odd charge NSC and even charge MIC, so that the former is vanished and the latter can be finite. Here we provide mirror reflection results, which can be viewed as a "directional" spatial inversion, aiding detailed discussion of different photocurrent components. In addition, the layer-dependent BPV current can also be analyzed here. For example, in the AF MBT-BT-MBT trilayer, the two MBT SLs are connected by $\mathcal{PT}$ ($\mathcal{PT}\mathbf{x}_{SL1} = \mathbf{x}_{SL2}$). Thus, the charge NSC and spin MIC in these SLs are flowing oppositely, while the charge MIC and spin NSC on them will be the same. Each MBT SL is $\mathcal{M}_x\mathcal{T}$, thus the $\sigma_{\text{NSC}}^x$, $\eta_{\text{MIC}}^y$, $\sigma_{\text{NSC}}^{x_z}$, and $\eta_{\text{MIC}}^{y_z}$ are all symmetrically forbidden.

Table II. LPL induced photocurrents under different symmetry operators. The mirror normal direction $i$ and $j$ are $x$ or $y$, so that the spin polarization $S_z$ always flip its sign under mirror operator.

|  | $\sigma_{\text{NSC}}^j$ | $\eta_{\text{MIC}}^j$ | $\sigma_{\text{NSC}}^{j_z}$ | $\eta_{\text{MIC}}^{j_z}$ |
| --- | --- | --- | --- | --- |
| $\mathcal{M}_i\ (i \neq j)$ | even | even | odd | odd |
| $\mathcal{M}_j$ | odd | odd | even | even |
| $\mathcal{P}$ | odd | odd | odd | odd |
| $\mathcal{T}$ | even | odd | odd | even |

One may wonder if out-of-plane BPV conductivity components of 2D materials can be similarly calculated and discussed. In the current theoretical framework, in-plane electric field component corresponds to normal incidence of light. Then the periodicity in the $xy$ plane eliminates the edge/end effects. According to previous work [68], this is akin to the closed circuit boundary condition, where electric field **E** can be used as the natural variable. As is well-known, in finite sized system, the edge/end effect could accumulate induced charges and drastically change the boundary condition. The out-of-plane electric field component corresponds to tangential incidence, which is experimentally challenging for thin films. In addition, the boundary condition becomes open circuit feature, and the charge accumulation on the film surface could be significant. In this situation, one should, in principle, adopt electric displacement **D** rather than **E** as the natural variable. On the other hand, the response function, formulated via Kubo perturbation theory [3], evaluates (charge and spin) current density. This definition is also in-plane, while it would be ill-defined for the out-of-plane current. Note that the $z$-direction is quantum confined for thin films, hence it is challenging to define a current along $z$. Recently, it is proposed that such out-of-plane response is actually electric dipole, and which requires evaluations of Wannier centers [69]. Nonetheless, the functions in the current work cannot be straightforwardly applied for out-of-plane component calculations.

We would like to remark that the magnetic direction would change the symmetry arguments since spin is a pseudovector. In this work, we only focus on the easy axis magnetic direction ($z$), which preserves the $C_{3z}$ rotation. If the magnetic moment is aligned in the $xy$-plane, or when paramagnetic configuration emerges, the $\mathcal{M}_x\mathcal{T}$ symmetry breaks, so that the photocurrents would change their mechanisms and flowing direction due to the reduction of its magnetic group. This is out of the scope of our current work, and will be discussed elsewhere.

**Conclusion.**

In the current work, we systematically calculate the layer-dependent photocurrent responses of MBT and BT superlattice thin films. Various sources of charge and spin BPV effects have been evaluated, and we provide symmetry constraints and arguments in detail. We show that even though a nonmagnetic BT QL is spatially centrosymmetric, one can break $\mathcal{P}$ via applying magnetic proximity effect through the vdW interfaces. In this regard, large charge and spin BPV currents can be generated. The direction of charge and spin current are vertical to each other, yielding pure charge and spin BPV photocurrent simultaneously. This strategy is different from previously proposed $\mathcal{P}$-boken methods in otherwise nonmagnetic systems, such as introducing dopants, manipulating structures, or applying electric bias. The BPV charge and spin currents can be tuned by magnetic configuration, light polarization, and interfacial effects.

**Acknowledgments.** This work was supported by the National Natural Science Foundation of China (NSFC) under Grant Nos. 11974270 and 21903063. The computational resources from HPC platform of Xi'an Jiaotong University and Hefei advanced computing center are also acknowledged.